\documentclass[prd,superscriptaddress,showpacs,preprintnumbers,twocolumn,nofootinbib]{revtex4-1}

\usepackage{amsmath,amssymb}
\usepackage{graphicx}

\begin{document}

\title{Leptonic mixing, family symmetries, and neutrino phenomenology}
\author{I. de Medeiros Varzielas}
\email{ivo.de@udo.edu}
\affiliation{Departamento de F\'{\i}sica and Centro de F\'{\i}sica Te\'{o}rica
de Part\'{\i}culas, Instituto Superior T\'{e}cnico, Avenida Rovisco Pais, 1049-001 Lisboa, Portugal}
\affiliation{Fakult\"{a}t f\"{u}r Physik, Technische Universit\"{a}t Dortmund D-44221 Dortmund, Germany}
\author{R. Gonz\'{a}lez Felipe}
\email{gonzalez@cftp.ist.utl.pt}
\affiliation{Departamento de F\'{\i}sica and Centro de F\'{\i}sica Te\'{o}rica
de Part\'{\i}culas, Instituto Superior T\'{e}cnico, Avenida Rovisco Pais, 1049-001 Lisboa, Portugal}
\affiliation{Instituto Superior de Engenharia de Lisboa, Rua Conselheiro Em\'{\i}dio Navarro, 1959-007 Lisboa, Portugal}
\author{H. Ser\^{o}dio}
\email{hserodio@cftp.ist.utl.pt}
\affiliation{Departamento de F\'{\i}sica and Centro de F\'{\i}sica Te\'{o}rica de Part\'{\i}culas, Instituto Superior T\'{e}cnico, Avenida Rovisco Pais, 1049-001 Lisboa, Portugal}

\begin{abstract}
Tribimaximal leptonic mixing is a mass-independent mixing scheme consistent with the present solar and atmospheric neutrino data. By conveniently decomposing the effective neutrino mass matrix associated to it, we derive generic predictions in terms of the parameters governing the neutrino masses. We extend this phenomenological analysis to other mass-independent mixing schemes which are related to the tribimaximal form by a unitary transformation. We classify models that produce tribimaximal leptonic mixing through the group structure of their family symmetries in order to point out that there is often a direct connection between the group structure and the phenomenological analysis. The type of seesaw mechanism responsible for neutrino masses plays a role here, as it restricts the choices of family representations and affects the viability of leptogenesis. We also present a recipe to generalize a given tribimaximal model to an associated model with a different mass-independent mixing scheme, which preserves the connection between the group structure and phenomenology as in the original model. This procedure is explicitly illustrated by constructing toy models with the transpose tribimaximal, bimaximal, golden ratio, and hexagonal leptonic mixing patterns.
\end{abstract}
\maketitle

\section{Introduction \label{Intro}}

Neutrino data~\cite{Fogli:2005cq,Schwetz:2008er,GonzalezGarcia:2010er} is well in agreement with tribimaximal (TB) mixing~\cite{Harrison:2002er}. If one assumes that the leptonic mixing is described at leading order by TB mixing, it is natural to consider that this special structure arises due to a family symmetry. Although continuous groups have been used \cite{King:2005bj,deMedeirosVarzielas:2005ax}, discrete symmetries are particularly attractive for this purpose, with $A_4$ being especially popular and featured in several models of leptonic mixing~\cite{Altarelli:2010gt}.

From the phenomenological viewpoint, an attractive feature of the mass-independent mixing schemes (also referred to as form-diagonalizable schemes) is that they lead to predictive neutrino mass matrix structures which contain just five parameters (the three neutrino masses and the two Majorana phases), in the absence of Dirac-type $CP$ violation. The latter can then be directly related to neutrino observables such as the mass-squared differences, the absolute mass scale, and the effective mass parameter in neutrinoless double beta ($0\nu\beta\beta$) decays.

In this paper, we parametrize the general effective neutrino mass matrix diagonalized by TB mixing through three independent contributions which reflect the well-known magic and $\mu$-$\tau$ symmetries, as well as the democratic symmetry. We then connect some phenomenological features directly to the magnitude of the parameter controlling the democratic contribution. Furthermore, we show that this analysis is also valid for other mass-independent mixing schemes by expressing a given form-diagonalizable structure in terms of the TB mixing rotated by an appropriate unitary matrix. We also discuss how to obtain in a natural way the above three independent contributions  from discrete non-Abelian symmetries. An alternative implementation of TB mixing in minimal $A_4$ (type-I) seesaw models based on the so-called form dominance can be found in Ref.~\cite{Chen:2009um}. Much of what applies to model building with TB mixing also transfers to other mass-independent schemes. To illustrate this, we build explicitly toy models that exhibit transpose tribimaximal~\cite{Fritzsch:1995dj}, bimaximal~\cite{Barger:1998ta}, golden ratio~\cite{Kajiyama:2007gx,Rodejohann:2008ir} and hexagonal~\cite{Giunti:2002sr} mixing patterns.

\section{Effective neutrino phenomenology}
\label{Pheno}

\subsection{Tribimaximal mixing phenomenology}
\label{TBP}

The effective neutrino mass matrix with TB mixing can be written without loss of generality in the form
\begin{align}\label{mnuTB}
\begin{split}
&m_\text{TB}=U_\text{TB} \, d_\nu  \, U_\text{TB}^T\\
&= \frac{1}{3}
{\small
\begin{pmatrix}
2x^\prime+3y^\prime+z^\prime&-x^\prime+z^\prime&-x^\prime+z^\prime\\
-x^\prime+z^\prime&2x^\prime+z^\prime&-x^\prime+3y^\prime+z^\prime\\
-x^\prime+z^\prime&-x^\prime+3y^\prime+z^\prime&2x^\prime+z^\prime
\end{pmatrix}}\,,
\end{split}
\end{align}
where $d_\nu=\text{diag}(x^\prime+y^\prime,y^\prime+z^\prime,x^\prime-y^\prime)$, and
\begin{align} \label{UTB}
U_\text{TB}=\begin{pmatrix}
\sqrt{\frac{2}{3}}&\frac{1}{\sqrt{3}}&0\\
-\frac{1}{\sqrt{6}}&\frac{1}{\sqrt{3}}&-\frac{1}{\sqrt{2}}\\
-\frac{1}{\sqrt{6}}&\frac{1}{\sqrt{3}}&\frac{1}{\sqrt{2}}
\end{pmatrix}
\end{align}
is the TB mixing matrix with angles $\theta_{12} = \arcsin (1/\sqrt{3})$, $\theta_{23}= -\pi/4$, and $\theta_{13}=0$, in the standard parametrization adopted in Ref.~\cite{Nakamura:2010zzi}.
The mass matrix $m_\text{TB}$ can be separated into three components,
\begin{align}\label{mnu3}
m_\text{TB}&=x^\prime C +y^\prime P+z^\prime D,
\end{align}
where
\begin{align}\label{CPDstruct}
C=\frac{1}{3}{\small\begin{pmatrix}
2&-1&-1\\
-1&2&-1\\
-1&-1&2
\end{pmatrix}}, P={\small \begin{pmatrix}
1&0&0\\
0&0&1\\
0&1&0
\end{pmatrix}}, D=\frac{1}{3}{\small
\begin{pmatrix}
1&1&1\\
1&1&1\\
1&1&1
\end{pmatrix}}
\end{align}
denote the well-known magic, $\mu$-$\tau$ symmetric, and democratic matrices, respectively.

As it turns out, through the above decomposition, we are able to reveal interesting phenomenological properties that apply to many models, classified according to the parameters $x^\prime$, $y^\prime$, and, particularly, $z^\prime$. For instance, if either $x^\prime$ or $y^\prime$ vanishes, the neutrino mass spectrum has a twofold degeneracy [see Eq.~\eqref{numasses} below]. Such a possibility is already excluded by the experimental data. Thus, only the contribution proportional to $z^\prime$, i.e., the democratic contribution, can be absent in Eq.~\eqref{mnu3}.

In the physical basis, where the charged leptons are diagonal and real, the effective low-energy leptonic mixing can be written in the form
\begin{align}
U_\nu=e^{-i \sigma_1/2}\,U_\text{TB}\,
\begin{pmatrix}
1&&\\
&e^{i\gamma_1}&\\
&&e^{i\gamma_2}
\end{pmatrix},
\end{align}
with $\gamma_1=(\sigma_1-\sigma_2)/2,\,\gamma_2=(\sigma_1-\sigma_3)/2$, being the Majorana phases and $\sigma_{1,3}=\text{arg}(x^\prime \pm y^\prime),\,\sigma_2=\text{arg}(y^\prime+z^\prime)$. In turn, the neutrino masses read as
\begin{align}\label{numasses}
\begin{split}
m_1&=\left|xe^{i\alpha_1}+y\right|=\left(x^2+y^2+2xy\cos\alpha_1\right)^{1/2},
\\
m_2&=\left|y+ze^{i\alpha_2}\right|=\left(y^2+z^2+2yz\cos\alpha_2\right)^{1/2},
\\
m_3&=\left|xe^{i\alpha_1}-y\right|=\left(x^2+y^2-2xy\cos\alpha_1\right)^{1/2},
\end{split}
\end{align}
where $x=|x^\prime|$, $y=|y^\prime|$, $z=|z^\prime|$ ,and
$\alpha_1=\text{arg}\,x^\prime-\text{arg}\,y^\prime$,
$\alpha_2=\text{arg}\,z^\prime-\text{arg}\,y^\prime$.

Experimentally, only the mass-squared differences have been directly measured. The solar and atmospheric mass-squared differences at $2\,\sigma$ confidence level are presently~\cite{Schwetz:2008er}
\begin{align}
\begin{split}\label{experimental}
\Delta m^2_\text{sol}&\equiv \;m_2^2-m_1^2\;=
\left(7.25 - 8.11\right)\times 10^{-5}\,\text{eV}^2\,,\\
\Delta m^2_\text{atm}&\equiv\left|m_3^2-m_1^2\right|=\left(2.18 -
2.64\right)\times 10^{-3}\,
\text{eV}^2\,,
\end{split}
\end{align}
with the best-fit values $\Delta m^2_\text{sol}=7.65\times10^{-5}\,\text{eV}^2$ and $\Delta m^2_\text{atm}=2.40\times10^{-3}\,\text{eV}^2$. There is also an indirect cosmological bound for the sum of the neutrino masses, $\sum m_i < 0.58$~eV~\cite{Komatsu:2010fb}.

Another important constraint can be obtained from $0\nu\beta\beta$ experiments, which access the quantity
\begin{equation}
m_{ee} =\left|\sum_{n=1}^3 (U_{\nu})_{1i}^2 \, m_i\right| \,.
\end{equation}
Despite large uncertainties from poorly known nuclear matrix elements, present data sets an upper bound on $|m_{ee}|$ in the range 0.2 to 1 eV at 90\% C.L.~\cite{KlapdorKleingrothaus:2000sn,Arnaboldi:2008ds,Wolf:2008hf}. This will be improved in forthcoming experiments, with an expected sensitivity of about $10^{-2}$~eV~\cite{Aalseth:2004hb}.

We remark that the sign of the neutrino mass difference $(m_3 - m_2)$ is dictated by the ordering of the neutrino masses, being positive for normal ordering and negative for inverted ordering. In terms of the parameters $x, y, z$ and the phases $\alpha_{1,2}$, from Eq.~\eqref{numasses}, one obtains
\begin{align}\label{m2relations}
\begin{split}
\Delta
m^2_{21}&=z\left(z+2y\cos\alpha_2\right)-x\left(x+2y\cos\alpha_1\right)\,,\\
\Delta m_{31}^2&=-4xy\cos\alpha_1\,.
\end{split}
\end{align}
The above relations allow us to reduce the number of free parameters from five to three. Yet, even in this case, neutrino mass models are not very predictive. As we will discuss, the democratic component in Eq.~\eqref{mnu3} is naturally absent or suppressed in many flavor models. It is then of interest to study the latter cases and their phenomenological implications. When the democratic contribution is absent ($z=0$), the first relation in Eq.~\eqref{m2relations} becomes $\Delta m^2_{21}=-x\left(x+2y\cos\alpha_1\right)$. Since, by definition, $\Delta m^2_{21}>0$, then $\pi/2 <\alpha_1 <3\pi/2$. The second relation in Eq.~\eqref{m2relations} thus implies $\Delta m^2_{31}>0$, which enforces a normal hierarchy. We emphasize that this applies to all flavor models with mass-independent mixing schemes that do not contain in the effective neutrino mass matrix the democratic matrix $D$ or its rotated counterpart in other mixing schemes (cf. Sec.~\ref{OMI}).

\begin{figure*}[t]
\begin{tabular}{cc}
\includegraphics[width=8cm]{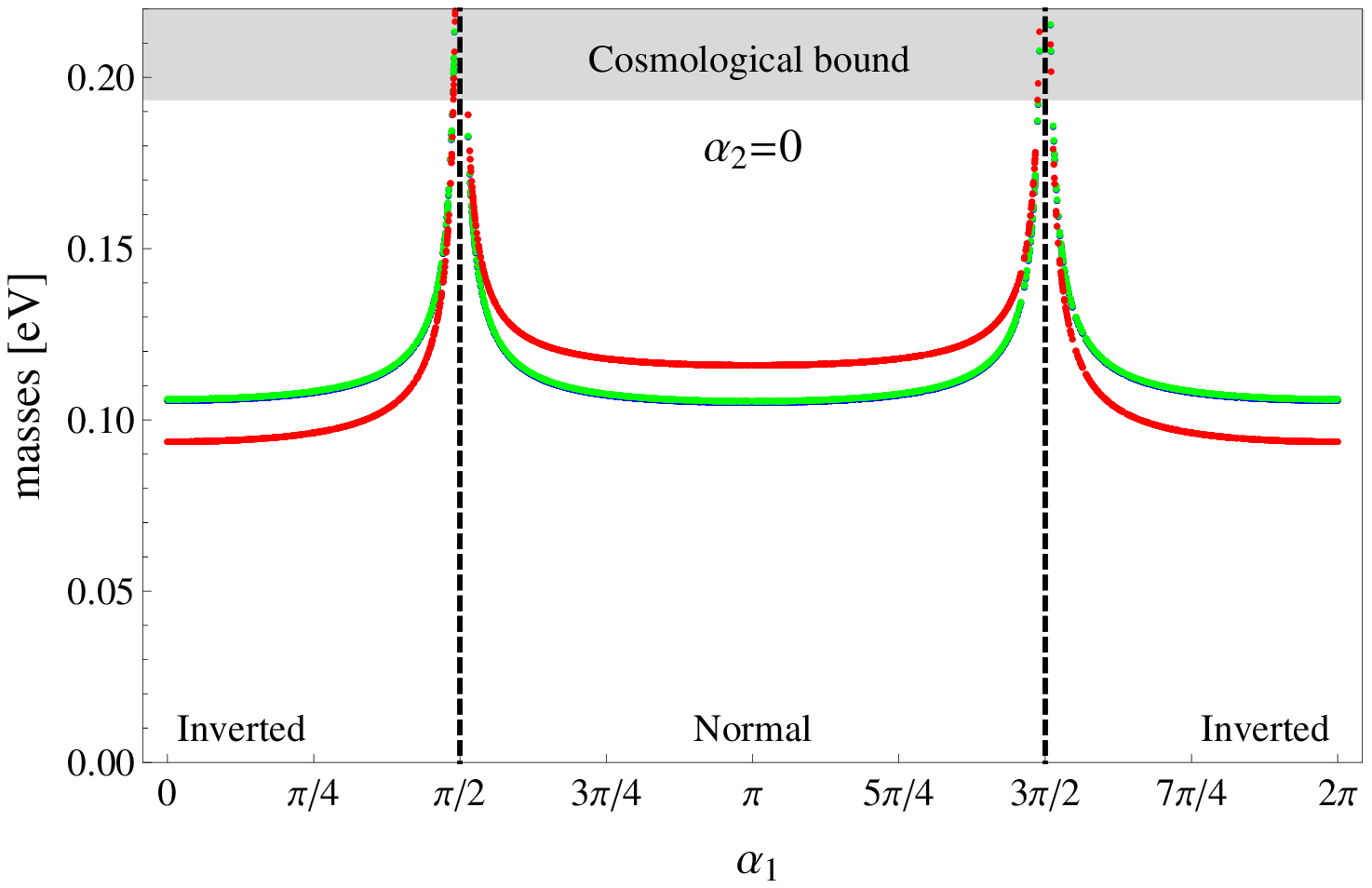}&
\includegraphics[width=7.6cm]{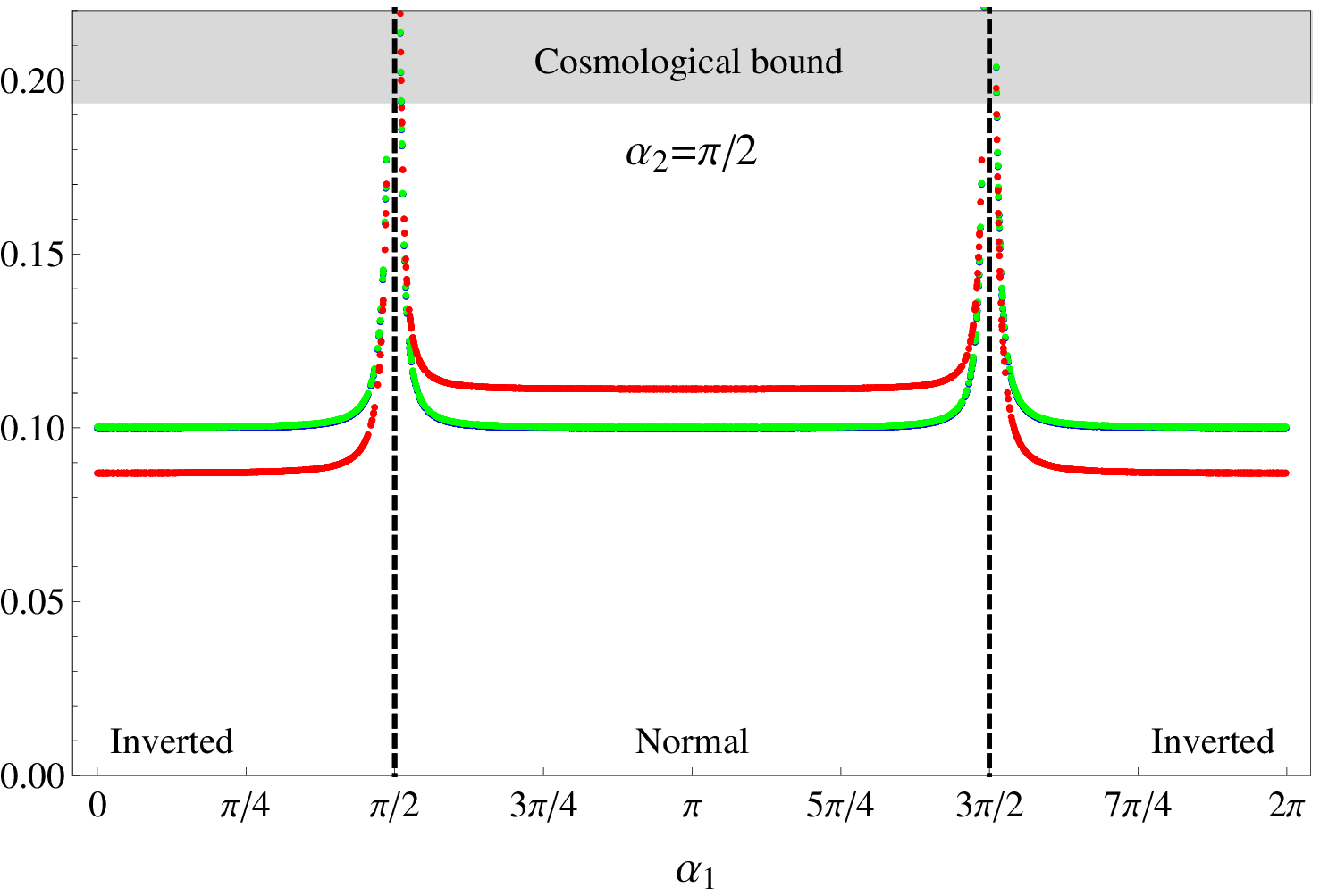}\\
\includegraphics[width=8cm]{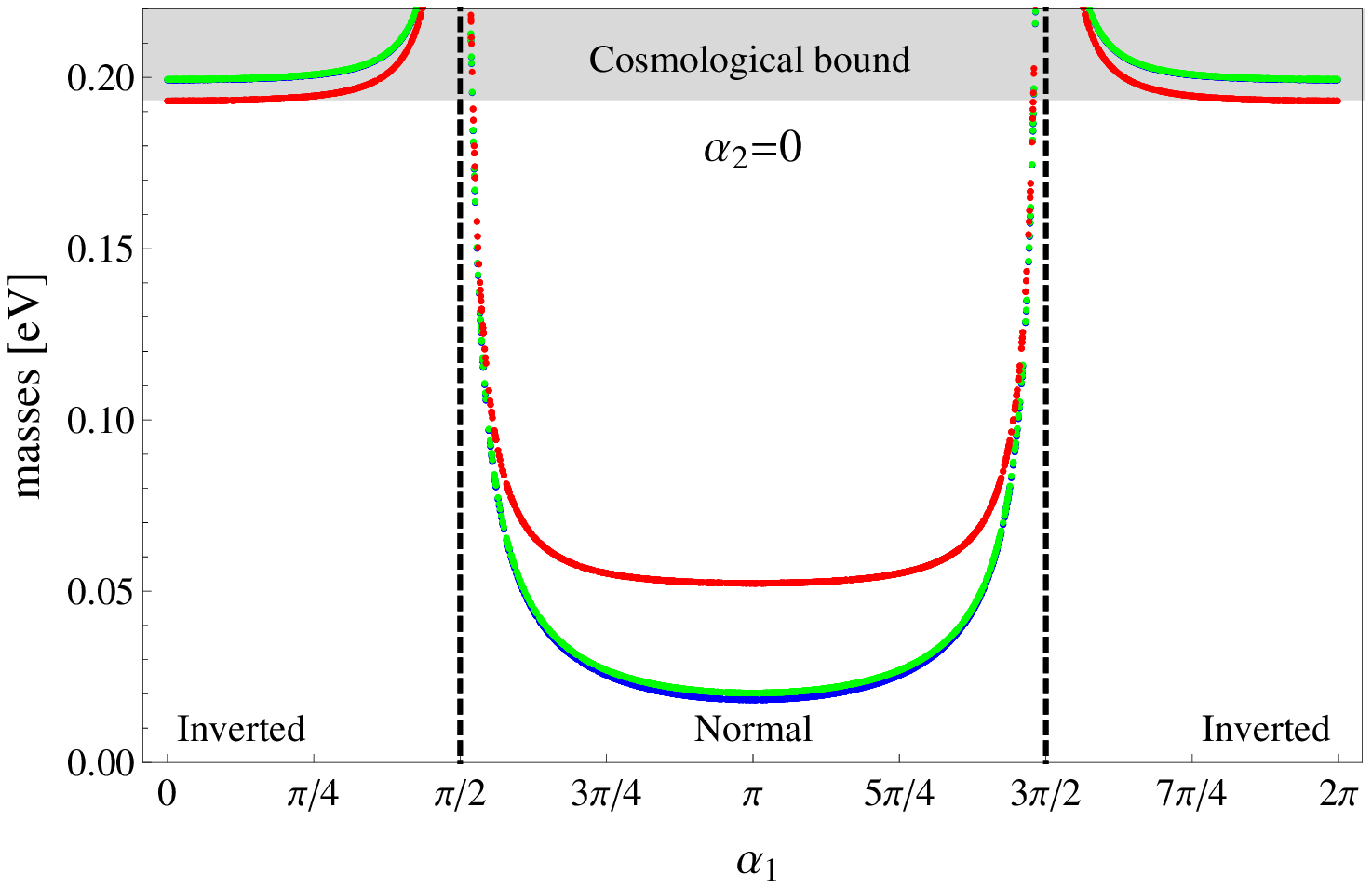}&
\includegraphics[width=7.6cm]{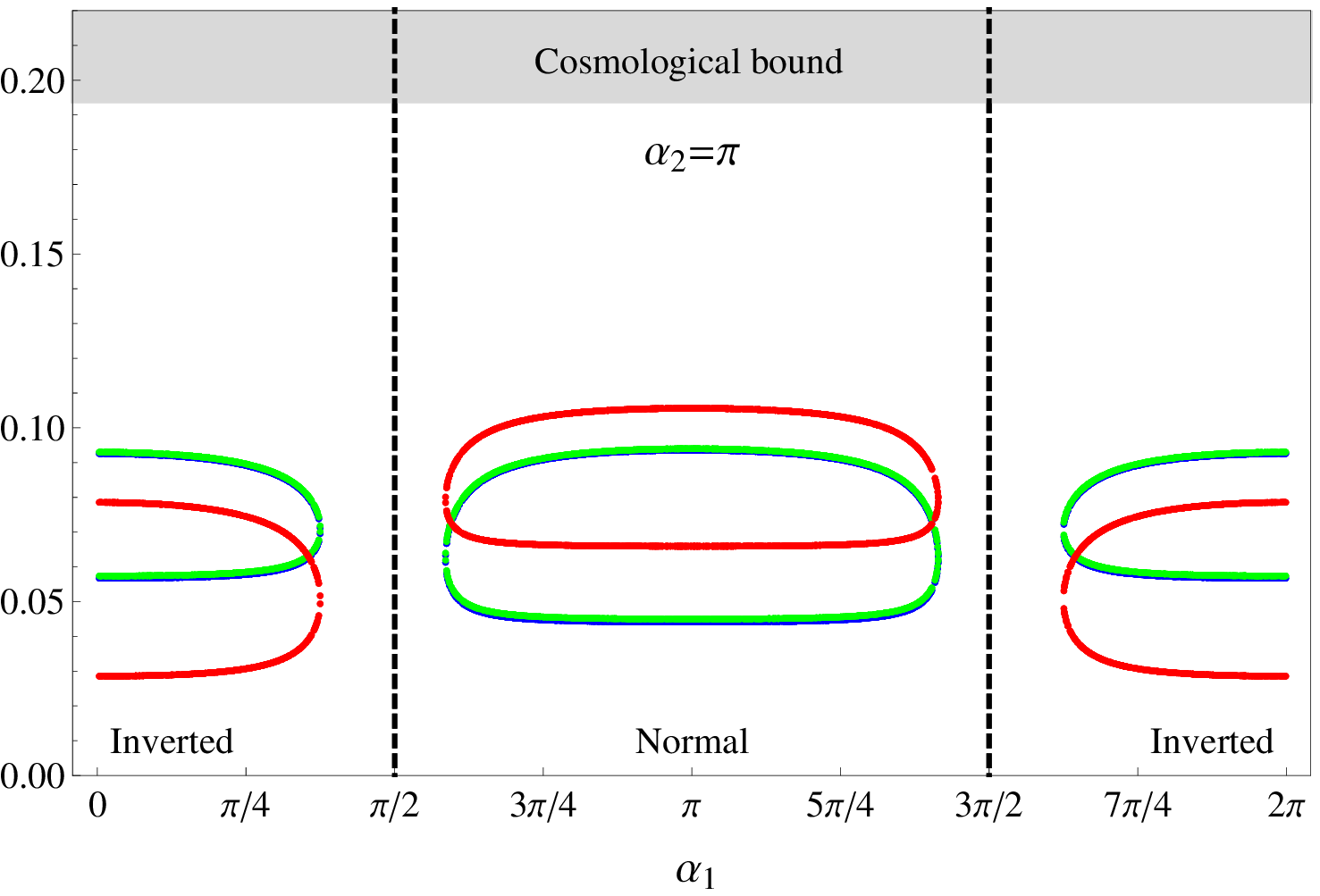}
\end{tabular}
\caption{\label{fig1}Neutrino mass spectrum as a function of the phase $\alpha_1$ for $z=0.1$~eV and $z=z_\text{lim} \simeq 3.3 \times 10^{-3}$~eV with $\alpha_2=0$ (left plots) and for $z=0.1$~eV and $\alpha_2=\pi/2, \pi$ (right plots). The curves correspond to the light neutrino masses $m_i$, properly ordered according to the spectrum hierarchy.}
\end{figure*}

Choosing, without loss of generality, $\alpha_1$ as the free parameter, the light neutrino masses become
\begin{align}\label{massesxy}
\begin{split}
m_1&=\left(y^2-\Delta m^2_{21}\right)^{1/2},\quad m_2=y,\\
m_3&=\left(y^2+\Delta m^2_{31}-\Delta m^2_{21}\right)^{1/2},
\end{split}
\end{align}
with
\begin{align}\label{xy}
\begin{split}
x&=\frac{\left(\Delta m^2_{31}-2\Delta m^2_{21}\right)^{1/2}}{\sqrt{2}}\,,\\
y&=-\frac{1}{2\sqrt{2}\cos\alpha_1}\frac{\Delta m^2_{31}}{\left(\Delta
m^2_{31}-2\Delta m^2_{21}\right)^{1/2}}\,.
\end{split}
\end{align}
The Majorana phases read $\gamma_1=\text{arg}(xe^{i\alpha_1}+y)/2$ and $\gamma_2=\gamma_1-\text{arg}(xe^{i\alpha_1}-y)/2$.

From Eqs.~\eqref{massesxy} and~\eqref{xy}, we conclude that the lightest neutrino mass has a lower bound, $m^\text{low}_1\simeq 1.56\times 10^{-2}$~eV, for $\alpha_1=\pi$. Moreover, the effective mass parameter $m_{ee}$ that governs $0\nu\beta\beta$ decay is approximately given by
\begin{equation}
\begin{split}
&m_{ee}=\frac{m_2}{3}\times\\
&\left[2\left(2-\frac{\Delta
m_{12}^2}{m_2^2}\right)(1+\cos{2\gamma_1})+1-2\frac{\Delta
m_{12}^2}{m_2^2}\right]^{1/2}
\end{split}
\end{equation}
and attains its lowest value $m_{ee}^\text{low}$ at $\alpha_1=\pi$, when $m_2$ is also minimal:
\begin{equation}
m_{ee}^\text{low}\simeq m_2^\text{low}\sqrt{1-\frac{2}{3}\frac{\Delta
m_{12}^2}{\left(m_2^\text{low}\right)^2}}\simeq 1.64\times 10^{-2}\,\text{eV}.
\end{equation}

\begin{figure*}[t]
\begin{tabular}{cc}
\includegraphics[width=8cm]{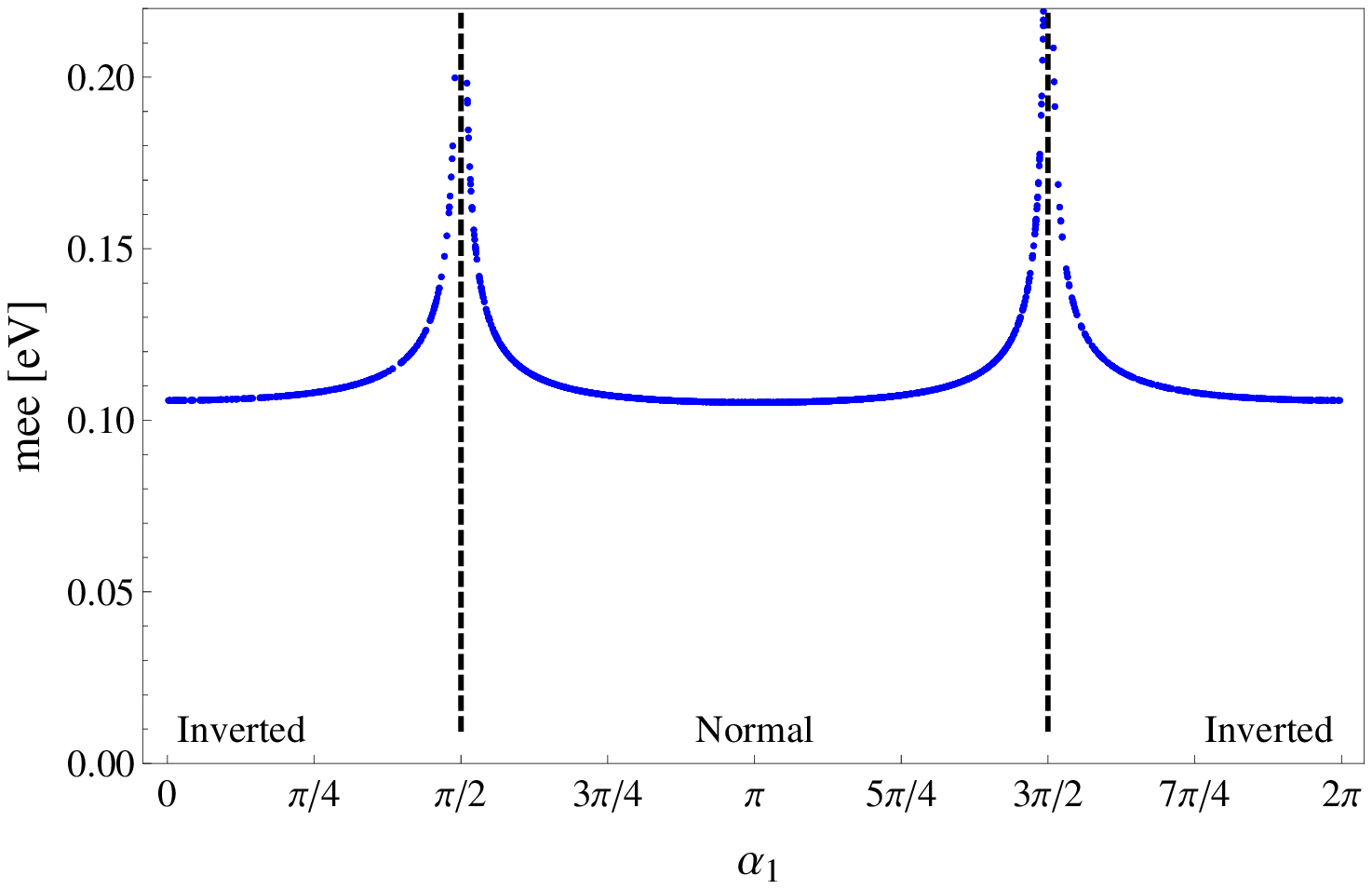}&
\includegraphics[width=7.6cm]{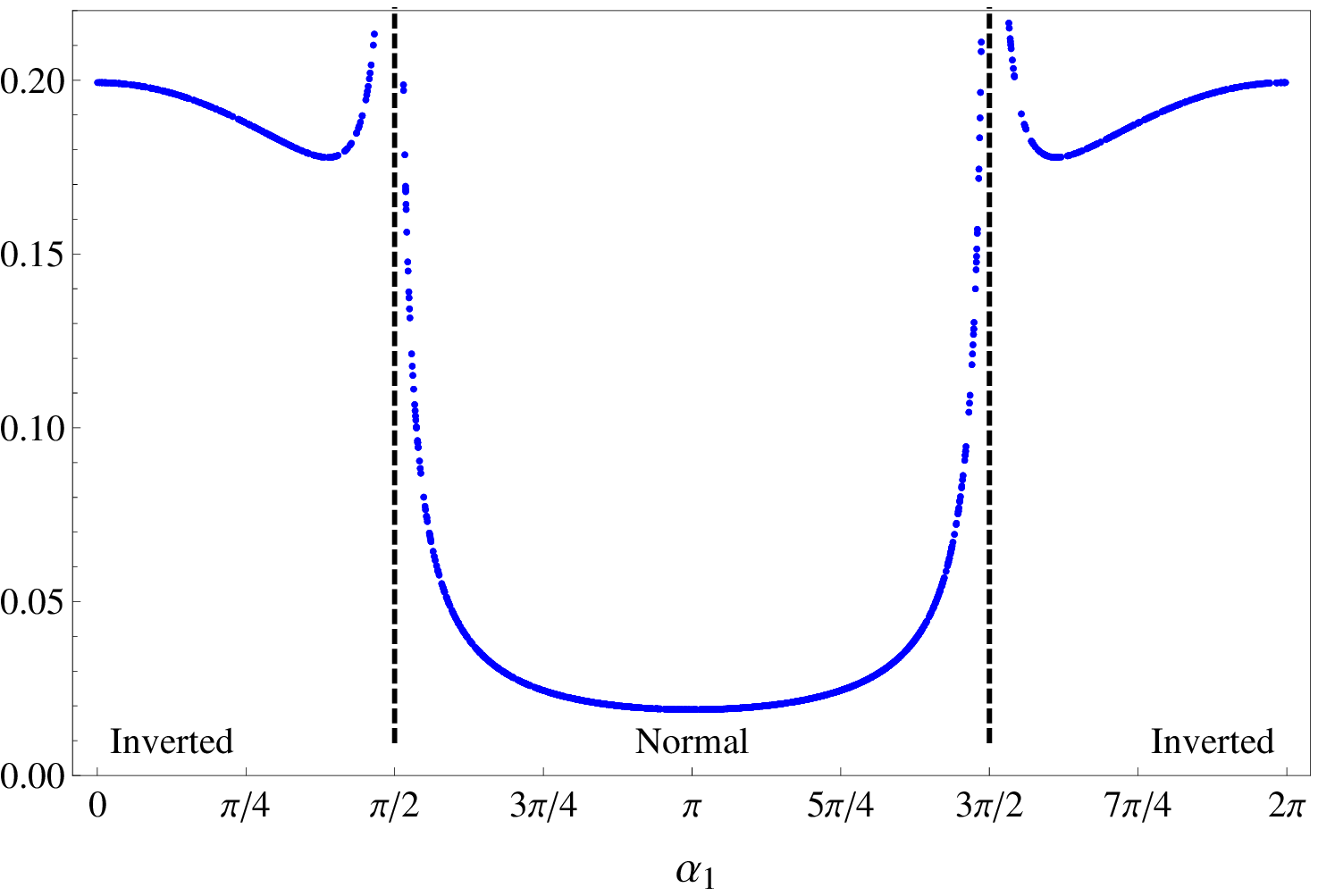}
\end{tabular}
\caption{\label{fig2}The effective Majorana mass parameter $m_{ee}$ for $\alpha_2=0$, assuming $z=0.1$~eV (left) and $z=z_\text{lim}\simeq 3.3 \times 10^{-3}$~eV (right).}
\end{figure*}

Let us now consider the case in which a small contribution from the democratic structure is present in the effective neutrino mass matrix. It can be seen from Eqs.~(\ref{m2relations}) that an inverted hierarchy is now allowed for small values of $z$. Furthermore, such a hierarchy is easier to achieve when $\alpha_2=0$; for other values of $\alpha_2$, the inverted hierarchy is, in general, excluded for $z \lesssim 0.01$~eV. Assuming small $z$ and $\alpha_2=0$, the mass spectrum is
\begin{align}
\begin{split}
m_1&\simeq\left(y^2-\Delta m_{21}^2+2yz\right)^{1/2},\\
m_2&\simeq\left(y^2+2yz\right)^{1/2},\\
m_3&\simeq\left(y^2+\Delta m_{31}^2-\Delta m_{21}^2+2yz\right)^{1/2},
\end{split}
\end{align}
so that the solar mass-squared difference in Eq.~\eqref{m2relations} can be approximated by $\Delta m_{21}^2 \simeq 2yz-x(x+2y\cos\alpha_1)$. Because of the presence of the $2yz$ term in this relation, an inverted neutrino spectrum is now viable. The parameters $x$ and $y$ no longer have a closed form, as in Eq.~\eqref{xy}. Nevertheless, one can solve numerically the equations for the mass spectrum. For illustration, in Fig.~\ref{fig1}, we present the neutrino mass spectrum as a function of the phase $\alpha_1$ for $z=0.1$~eV and $\alpha_2=0,\, \pi/2,\, \pi$. We find that, for $z \ge z_\text{lim} \simeq 3.3 \times 10^{-3}$~eV, an inverted mass hierarchy is allowed. The limiting case is depicted in the lower left plot of Fig.~\ref{fig1}, which corresponds to the limiting value $z=z_\text{lim}$ and a vanishing phase $\alpha_2$. In Fig.~\ref{fig2}, we present the $0\nu\beta\beta$ parameter $m_{ee}$ for $\alpha_2=0$ and two different values of the democratic contribution to the neutrino mass matrix, $z=0.1$~eV and $z=z_\text{lim}$.

\subsection{Other mass-independent mixing schemes \label{OMI}}

In addition to TB mixing, there are other mass-independent structures that can reproduce the observed leptonic mixing angles. Below we give some examples of such mass-independent schemes.

The transposed TB mixing has the mixing angles $\theta_{12} = \pi/4$, $\theta_{23}= \arctan \sqrt{2}$, and $\theta_{13}=0$~\cite{Fritzsch:1995dj}. In the well-known bimaximal structure~\cite{Barger:1998ta}, the mixing angles are $\theta_{12} = \theta_{23} = \pi/4$ and $\theta_{13}=0$. There are also two golden ratio (GR) proposals where the angles can be related with $\Phi=(1+\sqrt{5})/2$. The first (GR1) scheme corresponds to $\theta_{12} = \arctan \left(1/\Phi\right)$, $\theta_{23}=\pi/4$, and $\theta_{13}=0$~\cite{Kajiyama:2007gx}, while the second (GR2) has the associated angles $\theta_{12} = \arccos \left(\Phi/2\right)$, $\theta_{23}=-\pi/4$, and $\theta_{13}=0$~\cite{Rodejohann:2008ir}. Finally, the so-called hexagonal mixing~\cite{Giunti:2002sr} is described by the angles $\theta_{12} = \pi/6$, $\theta_{23}=-\pi/4$, and $\theta_{13}=0$.

The phenomenological analysis previously discussed is straightforwardly generalized to other mass-independent structures such as the five examples given above. Indeed, if the effective neutrino mass matrix $m_\nu$ is exactly diagonalized by the unitary matrix $U_X$ in a given mass-independent mixing scheme, one has
\begin{equation}
m_\nu = U_{X}\, d_\nu\, U_{X}^T.
\end{equation}
Reexpressing the mixing in terms of $U_\text{TB}$ and an appropriate rotation $K_X$,
\begin{equation}\label{KX}
U_X = K_X\, U_\text{TB},
\end{equation}
we can then rewrite
\begin{equation}
m_\nu = K_{X}\, m_\text{TB}\, K_{X}^T.
\end{equation}

We can see that the decomposition of the neutrino mass matrix into three independent components, as given in Eq.~\eqref{mnu3}, is maintained as each component is rotated appropriately by $K_X$. Thus, the analysis in terms of the parameters $x', y', z'$ also holds. From this simple manipulation, it should become clear that, although the mixing matrix is different, the main conclusions drawn for the TB mixing remain valid for any other mass-independent mixing structure, as well. In particular, the results that depend only on the neutrino mass spectrum are unchanged. On the other hand, since the effective mass parameter $m_{ee}$ depends directly on the first row of the mixing matrix, it is affected by the mixing scheme used. Yet, as the mixing angles are constrained by the experimentally allowed ranges, all the mass-independent schemes presented above lead to very similar predictions for $m_{ee}$.

\section{Leptonic mixing from family symmetry invariants \label{FSI}}

The $A_4 = \Delta(12)$ symmetry is often used to obtain TB mixing, due to its relative simplicity - it is the smallest discrete group with a triplet irreducible representation (irrep). As a function of the parameters $x', y'$, and $z'$ introduced in Eqs.~\eqref{mnuTB} and \eqref{mnu3}, the TB effective Lagrangian can be written as
\begin{align}
\label{TB_xyz_terms}
\begin{split}
\mathcal{L}_\nu \propto&\,
x^\prime \left(2 \nu_1 \nu_1 - \nu_1 \nu_2 - \nu_1 \nu_3 - \nu_2 \nu_1 + 2 \nu_2
\nu_2 -
\nu_2 \nu_3\right.\\
&\left. - \nu_3 \nu_1 - \nu_3 \nu_2 + 2 \nu_3 \nu_3\right)+y^\prime (\nu_1 \nu_1
+ \nu_2 \nu_3 + \nu_3 \nu_2)\\
&+z^\prime (\nu_1 \nu_1 + \nu_1 \nu_2 + \nu_1 \nu_3 + \nu_2 \nu_1 + \nu_2 \nu_2
+
\nu_2 \nu_3\\
& + \nu_3 \nu_1 + \nu_3 \nu_2 + \nu_3 \nu_3) \,.
\end{split}
\end{align}
In order to show how the Lagrangian structure of Eq.~(\ref{TB_xyz_terms}) arises in $A_4$ models, we reproduce below the most relevant group theoretical aspects (a more detailed study can be found, e.g., in Ref.~\cite{Luhn:2007uq}). We then extend our analysis to the $\Delta(3 n^2)$ family symmetry groups.

\subsection{$A_4$ invariants}

The $A_4$ group has a $Z_3$ and a $Z_2$ generator that are not simultaneously diagonalizable. It has three singlet irreps and one triplet irrep, the latter being denoted as $(0,1)$ in the notation used in Ref.~\cite{Luhn:2007uq} for $\Delta(3 n^2)$. As we shall see below, the triplet representation is labeled according to the $Z_n$ generators, in the basis where they are diagonal. Some of the invariants that can be constructed are $(0,1) \times (0,1)$ and $(0,1) \times (0,1) \times (0,1)$.

In the basis where the $Z_3$ generator is diagonal, the $(0,1) \times (0,1)$ invariant becomes $\varphi^\prime_1\varphi_1+\varphi^\prime_2\varphi_3 +\varphi_3^\prime\varphi_2$, where $\varphi_i$ and $\varphi^\prime_i$ are the components of the two triplets. We then immediately conclude from the group structure that the required effective neutrino mass term $c_{ij} L_i H L_j H$ can be constructed if the lepton doublet $L_i$ is assigned to a $(0,1)$ irrep. Here, $i$ is a generation index, and $H$ denotes the standard model (SM) Higgs field. Furthermore, adding a SM singlet $(0,1)$ scalar (familon) field $\phi$, the invariant term $L H L H \phi$ can also be constructed, which becomes a mass term when the familon acquires a vacuum expectation value (VEV).

In the following, we assume that, through the appropriate scalar potential, the aligned VEV $\langle \phi \rangle \propto (1,1,1)$ is obtained\footnote{In a complete model this is a nontrivial issue that should be addressed.}. Such a VEV alignment leads to both the $\mu$-$\tau$ and magic structures given in Eq.~\eqref{CPDstruct} and yields the Lagrangian of Eq.~(\ref{TB_xyz_terms}) with $z'=0$. In order to produce the democratic structure, one can use higher order terms\footnote{This approach is usually adopted in continuous models~\cite{King:2005bj,deMedeirosVarzielas:2005ax} and the so-called indirect
models~\cite{deMedeirosVarzielas:2006fc}.} such as $(L H \phi)(L H \phi)$. It is, therefore, legitimate to consider that, in models based on $A_4$, the parameter $z'$ in Eq.~(\ref{mnu3}) naturally vanishes or is small in comparison with other contributions.

\subsection{Generalization to $\Delta(3 n^2)$ \label{3nn}}

The group $\Delta(3 n^2) \sim \left(Z_n\times Z_n\right)\rtimes Z_3$ has a $Z_3$ generator  and two $Z_n$ generators, denoted here by $a$, $c$, and $d$, respectively. In the $Z_n$-diagonal basis, the generators of this group can be written in a triplet representation as follows:
\begin{equation}
a =
\begin{pmatrix}
0 & 1 & 0\\
0 & 0 & 1\\
1 & 0 & 0
\end{pmatrix},
\end{equation}
and
\begin{equation}
c =
\begin{pmatrix}
\eta^{k } & 0 & 0\\
0 & \eta^{l} & 0\\
0 & 0 & \eta^{-k-l}
\end{pmatrix}\,,\quad
d =
\begin{pmatrix}
\eta^{-k-l } & 0 & 0\\
0 & \eta^{l} & 0\\
0 & 0 & \eta^{k}
\end{pmatrix},
\end{equation}
where $\eta=\exp(2\pi i/n)$ and the integers $k,l=0,1,...,(n-1)$ label the triplet representations. Because of the particular form of the group generators, a triplet irrep can be labeled in three different ways,
\begin{equation}
\begin{pmatrix}
k\\
l
\end{pmatrix}\rightarrow
\begin{bmatrix}
k\\
l
\end{bmatrix}_p\equiv
M^p\begin{pmatrix}
k\\
l
\end{pmatrix},
\end{equation}
with $p=0,1$, or 2, and
\begin{equation}
M=\begin{pmatrix}
-1&-1\\
1&0
\end{pmatrix}.
\end{equation}
When building $\Delta(3 n^2)$ invariants, we need to consider two cases: $n\neq 3\mathbb{Z}$ or $n= 3\mathbb{Z}$. The way the relevant invariants, $(\mathbf{3}\otimes\mathbf{3})_{\mathbf{1}_0}$ and $(\mathbf{3}\otimes\mathbf{3}\otimes\mathbf{3})_{\mathbf{1}_0}$, are obtained will be shown next. Hereafter, we shall use the simplified notation $(k,l)$ to denote $\mathbf{3}_{(k,l)}$. Moreover, when presenting the invariant products in terms of components, we shall assume $\mathbf{3}\sim \varphi$, with $\varphi=(\varphi_1,\varphi_2,\varphi_3)$.

\subsubsection{$n\neq 3 \mathbb{Z}$}

In this case, there are three one-dimensional $(r=0,1,2)$ and $(n^2-1)/3$ three-dimensional irreducible representations. We are only interested in the singlet contributions
\begin{equation}
\left(\mathbf{3}_{(k^\prime,l^\prime)}\otimes\mathbf{3}_{(k,l)}\right)_{\mathbf
{1}}= \mathbf{1}_0+\mathbf{1}_1+\mathbf{1}_2 \,,
\end{equation}
which verify the condition
\begin{equation}\label{cond2triplet}
\begin{pmatrix}
k^\prime\\
l^\prime
\end{pmatrix}=
\begin{bmatrix}
-k\\
-l
\end{bmatrix}_p\,.
\end{equation}
Without ambiguity we choose $p=0$. The Clebsch-Gordan coefficients for the invariant product of two triplets is given by
\begin{equation}
\mathbf{1}^{2\varphi}_r:\quad \varphi^\prime
\begin{pmatrix}
1&0&0\\
0&\omega^{-1}&0\\
0&0&\omega
\end{pmatrix}^r\varphi\,,
\end{equation}
where $\omega=\exp(2\pi i/3)$.
For the product of three triplets,
$\mathbf{3}_{(k^{\prime\prime},l^{\prime\prime})}
\otimes\mathbf{3}_{(k^{\prime},l^{\prime})}
\otimes\mathbf{3}_{(k,l)}$, the necessary condition for the invariant is
\begin{equation}\label{tricond}
\begin{pmatrix}
k^{\prime\prime}\\
l^{\prime\prime}
\end{pmatrix}=
\left[
\begin{pmatrix}
k^{\prime}\\
l^{\prime}
\end{pmatrix}+
\begin{bmatrix}
k\\
l
\end{bmatrix}_q
\right]_p ,
\end{equation}
and the Clebsch-Gordan coefficients are
\begin{align}
\mathbf{1}^{3\varphi}_r:\quad \varphi^{\prime\prime}a^p
\begin{pmatrix}
\varphi^\prime_1&0&0\\
0&\omega^{-r}\varphi^\prime_2&0\\
0&0&\omega^r\varphi^\prime_3
\end{pmatrix} a^q\varphi\,.
\end{align}
The variable $q=0,1,2$ is associated with one of the three triplets coming out of the product of two triplets, i.e.,
\begin{align}
\begin{split}
q=0:&\,\begin{bmatrix}k^\prime+k\\l^\prime+l\end{bmatrix}_p,\quad
q=1:\,\begin{bmatrix}k^\prime-k-l\\l^\prime+k\end{bmatrix}_p,\\
q=2:&\,\begin{bmatrix}k^\prime+l\\l^\prime-k-l\end{bmatrix}_p,
\end{split}
\end{align}
while the variable $p$ is either 0, 1, or 2. When the condition of Eq.~\eqref{cond2triplet} is satisfied, the triplet with $q=0$ is not present.

We are particularly interested in products of two and three triplets where two of them are the same. There are two different possibilities when $(k^\prime,l^\prime)=(k,l)$:

\begin{itemize}
\item[(i)] Case $(k,l)=(-k,-l)$

In this case, only triplets of the type $(0,n/2)$ and equivalent representations are allowed. From Eq.~\eqref{tricond}, we see that $q=0$ is not allowed so that the invariant products are
\begin{align}
\begin{split}
&\mathbf{1}^{2\varphi}_r\sim\begin{pmatrix}
0\\
n/2
\end{pmatrix}\times
\begin{pmatrix}
0\\
n/2
\end{pmatrix},\\
&\mathbf{1}^{3\varphi}_{r}\sim
\begin{pmatrix}
0\\
n/2
\end{pmatrix}\times
\begin{pmatrix}
0\\
n/2
\end{pmatrix}\times
\begin{pmatrix}
0\\
n/2
\end{pmatrix}\,,\,q=1,2.
\end{split}
\end{align}

\item[(ii)] Case $(k,l)\neq(-k,-l)$

It is not possible to write the invariant of two triplets in this case. However, there is one extra invariant in the product of three triplets:
\begin{align}
\begin{split}
&\mathbf{1}^{3\varphi}_r\sim\begin{pmatrix}
2k\\
2l
\end{pmatrix}\times
\begin{pmatrix}
k\\
l
\end{pmatrix}\times
\begin{pmatrix}
k\\
l
\end{pmatrix}\,,\,q=0,\\
&\mathbf{1}^{3\varphi}_r\sim\begin{pmatrix}
-k\\
-l
\end{pmatrix}\times
\begin{pmatrix}
k\\
l
\end{pmatrix}\times
\begin{pmatrix}
k\\
l
\end{pmatrix}\,,\,q=1,2.
\end{split}
\end{align}
\end{itemize}
Notice that, in both cases, we have assumed $p=q$, since these are the relevant invariants that will allow us to obtain TB mixing with a simple VEV alignment.

\subsubsection{$n=3\mathbb{Z}$}
When $n$ is an integer multiple of three, there exist nine one-dimensional $(s,r=0,1,2)$ and $(n^2-3)/3$ three-dimensional irreducible representations. Once again, we shall be interested in the singlet contributions. These are
\begin{align}
\left(\mathbf{3}_{(k^\prime,l^\prime)}\otimes\mathbf{3}_{(k,l)}\right)_{\mathbf{
1}}=
\sum_{s=0}^2 \left(\mathbf{1}_{0,s}+\mathbf{1}_{1,s}+\mathbf{1}_{2,s}\right)\,,
\end{align}
with the condition
\begin{align}
\begin{pmatrix}k^\prime\\l^\prime\end{pmatrix} = \begin{bmatrix}
-k+sn/3\\-l+sn/3\end{bmatrix}_p.
\end{align}

We have two different possibilities: $k^\prime$, $l^\prime$, $k$, and $l$ are either all multiples of $n/3$, or not all of them are multiples of $n/3$. In the former case, the condition for an invariant of the product of two triplets is
\begin{equation}
\begin{pmatrix}
-k^\prime\\
-l^\prime
\end{pmatrix}=
\begin{pmatrix}
k\\
l
\end{pmatrix}=
\begin{pmatrix}
0\\
\pm n/3
\end{pmatrix},
\end{equation}
with the Clebsch-Gordan coefficients given by
\begin{align}
\mathbf{1}_{r,s}^{2\varphi}:\quad \varphi^\prime a^{\mp s}
\begin{pmatrix}
1&0&0\\
0&\omega^{-1}&0\\
0&0&\omega
\end{pmatrix}^r\varphi \,.
\end{align}
For the invariant constructed from three triplets, the necessary condition is
\begin{equation}
\begin{pmatrix}
k^{\prime\prime}\\
l^{\prime\prime}
\end{pmatrix}=
\begin{pmatrix}
k^\prime\\
l^\prime
\end{pmatrix}=
\begin{pmatrix}
k\\
l
\end{pmatrix}=
\begin{pmatrix}
0\\
\pm n/3
\end{pmatrix},
\end{equation}
and the Clebsch-Gordan coefficients are
\begin{equation}
\mathbf{1}_{r,s}^{3\varphi}:\quad \varphi^{\prime\prime}a^{\pm s}a^q
\begin{pmatrix}
\varphi^\prime_1&0&0\\
0&\omega^{-r}\varphi^\prime_2&0\\
0&0&\omega^r \varphi^\prime_3
\end{pmatrix}a^q\varphi \,.
\end{equation}
The triplets associated with $q$ are the same as in the previous case. It is not possible to build an invariant from the product of two triplets when they are in the same irrep. On the other hand, for the product of three triplets, we get
\begin{equation}
\mathbf{1}_{r,s}^{3\varphi}:\quad
\begin{pmatrix}
0\\
\pm n/3
\end{pmatrix}\times
\begin{pmatrix}
0\\
\pm n/3
\end{pmatrix}\times
\begin{pmatrix}
0\\
\pm n/3
\end{pmatrix}\,,\,q=0,1,2.
\end{equation}

The second case, i.e., when not all $k^\prime$, $l^\prime$, $k$, and $l$ are multiples of $n/3$, is similar to the case with $n\neq 3\mathbb{Z}$. The necessary condition for the invariant product of two triplets is
\begin{equation}
\begin{pmatrix}
k^\prime\\
l^\prime
\end{pmatrix}=
\begin{bmatrix}
-k\\
-l
\end{bmatrix}_p+
\frac{s\,n}{3}
\begin{pmatrix}
1\\
1
\end{pmatrix}\,,
\end{equation}
with the Clebsch-Gordan coefficients of the form
\begin{equation}
\mathbf{1}_{r,s}^{2\varphi}:\quad \varphi^\prime s^p
\begin{pmatrix}
1&0&0\\
0&\omega^{-1}&0\\
0&0&\omega
\end{pmatrix}^r\varphi.
\end{equation}
The condition for the product of three triplets is
\begin{equation}
\begin{pmatrix}
k^\prime\\
l^\prime
\end{pmatrix}=\left[
-\begin{pmatrix}
k^\prime\\
l^\prime
\end{pmatrix}-
\begin{bmatrix}
k\\
l
\end{bmatrix}_q
\right]_p+\frac{s\,n}{3}
\begin{pmatrix}
1\\
1
\end{pmatrix}\,,
\end{equation}
and the corresponding Clebsch-Gordan coefficients read
\begin{equation}
\mathbf{1}_{r,s}^{3\varphi}:\quad \varphi^{\prime\prime}a^p
\begin{pmatrix}
\varphi_1^\prime&0&0\\
0&\omega^{-r}\varphi_2^\prime&0\\
0&0&\omega\varphi_1^\prime
\end{pmatrix}a^q\varphi.
\end{equation}
As in the case of $n \neq 3\mathbb{Z}$, we are only interested in the situation when $s=0$. We consider again two different possibilities when $(k^\prime,l^\prime)=(k,l)$.

\begin{itemize}
\item[(i)]  Case $(k,l)=(-k,-l)$

If $n$ is odd, it is not possible to satisfy this condition. For $n$ even we have $(k,l)=(0,n/2)$. Since the value $q=0$ is not allowed, we get
\begin{align}
\begin{split}
&\mathbf{1}^{2\varphi}_{r,0}\sim\begin{pmatrix}
0\\
n/2
\end{pmatrix}\times
\begin{pmatrix}
0\\
n/2
\end{pmatrix},\\
&\mathbf{1}^{3\varphi}_{r,0}\sim
\begin{pmatrix}
0\\
n/2
\end{pmatrix}\times
\begin{pmatrix}
0\\
n/2
\end{pmatrix}\times
\begin{pmatrix}
0\\
n/2
\end{pmatrix}\,,\,q=1,2.
\end{split}
\end{align}

\item[(ii)] Case $(k,l)\neq(-k,-l)$

In this case it is not possible to write the invariant of two triplets. The extra invariant in the product of three triplets is now given by
\begin{align}
\begin{split}
&\mathbf{1}^{3\varphi}_{r,0}\sim\begin{pmatrix}
-2k\\
-2l
\end{pmatrix}\times
\begin{pmatrix}
k\\
l
\end{pmatrix}\times
\begin{pmatrix}
k\\
l
\end{pmatrix}\,,\,q=0,\\
&\mathbf{1}^{3\varphi}_{r,0}\sim\begin{pmatrix}
k\\
l
\end{pmatrix}\times
\begin{pmatrix}
k\\
l
\end{pmatrix}\times
\begin{pmatrix}
k\\
l
\end{pmatrix}\,,\,q=1,2.
\end{split}
\end{align}

\end{itemize}

\subsubsection{Invariant decomposition}

In the previous section, we presented a simple approach to build the invariants of the $\Delta(3n^2)$ group.  We also constructed the irrep combinations and their associated Clebsch-Gordan decomposition for the relevant invariants. The results were obtained in the basis where the $Z_n$ generators are diagonal. Yet, for model building, it is more convenient to change to the basis where the $Z_3$ generator is diagonal. The only transformation that is needed in order to get the Clebsch-Gordan decomposition in the new basis is given by
\begin{align}\label{UW}
\varphi = U(\omega)^\dagger\, \varphi_\text{old} \,,\quad
U(\omega)=\frac{1}{\sqrt{3}}
\begin{pmatrix}
1&1&1\\
1&\omega&\omega^2\\
1&\omega^2&\omega
\end{pmatrix}\,,
\end{align}
where $\varphi_\text{old}$ denotes the triplet field in the old basis. In the new basis, the invariant for the product of two triplets is
\begin{equation}
\mathbf{1}^{2\varphi}_{0(0)}
:\quad\varphi^\prime_1\varphi_1+\varphi^\prime_2\varphi_3
+\varphi_3^\prime\varphi_2\,,
\end{equation}
while for the product of three triplets the invariants are
\begin{widetext}
\begin{align}
\begin{split}
\mathbf{1}^{3\varphi}_{0(0)} (q=0):&\quad
\frac{1}{\sqrt{3}}\left\{\varphi^{\prime\prime}_1(\varphi^\prime_1\varphi_1
+\varphi^\prime_2\varphi_3+\varphi^\prime_3\varphi_2)+\varphi^{\prime\prime}
_2(\varphi^\prime_2\varphi_2
+\varphi^\prime_1\varphi_3+\varphi^\prime_3\varphi_1)+\varphi^{\prime\prime}
_3(\varphi^\prime_3\varphi_3
+\varphi^\prime_1\varphi_2+\varphi^\prime_2\varphi_1)\right\},
\\\\
\mathbf{1}^{3\varphi}_{0(0)} (\text{sym}):&\quad
\frac{1}{\sqrt{3}}\left\{\varphi^{\prime\prime}_1(2\varphi^\prime_1\varphi_1
-\varphi^\prime_2\varphi_3-\varphi^\prime_3\varphi_2)+\varphi^{\prime\prime}
_2(2\varphi^\prime_2\varphi_2
-\varphi^\prime_1\varphi_3-\varphi^\prime_3\varphi_1)+\varphi^{\prime\prime}
_3(2\varphi^\prime_3\varphi_3
-\varphi^\prime_1\varphi_2-\varphi^\prime_2\varphi_1)\right\},\\\\
\mathbf{1}^{3\varphi}_{0(0)} (\text{asym}):&
\quad
i\left(\varphi^{\prime\prime}_1\varphi^\prime_2\varphi_3+\varphi^{\prime\prime}
_2\varphi^\prime_3\varphi_1
+\varphi^{\prime\prime}_3\varphi^\prime_1\varphi_2-\varphi^{\prime\prime}
_1\varphi^\prime_3\varphi_2-\varphi^{\prime\prime}
_2\varphi^\prime_1\varphi_2-\varphi^{\prime\prime}
_3\varphi^\prime_2\varphi_1\right) \,.
\end{split}
\end{align}
\end{widetext}

In this basis, the charged lepton sector is diagonal and TB mixing is easily achieved in the neutrino sector from the products $\mathbf{3} \otimes \mathbf{3}$ and $\mathbf{3} \otimes \mathbf{3} \otimes \mathbf{3}$ with simple VEVs.

Although the Clebsch-Gordan decomposition for other mass-independent mixing schemes is comparably more complicated in the new basis, it can be obtained through the rotation matrix $K_X$, given in Eq.~\eqref{KX}. Moreover, the change to the convenient $Z_3$-diagonal basis is performed analogously to the one in Eq.~\eqref{UW}, where $U(\omega)$ is now replaced by $U(\omega) K_X^\dagger$. This is a good basis choice to obtain other mass-independent mixing patterns, since the rotated $\mathbf{3} \otimes \mathbf{3}$ directly leads to the rotated $\mu$-$\tau$ symmetric matrix $P$, as in the TB mixing case.

\section{Types of seesaw}

Under general considerations, the effective neutrino mass term is written as the nonrenormalizable operator $L_i H L_j H (...)$, where $(...)$ denotes additional fields that may be present and increase the dimensionality of the operator. To establish what kind of family invariants lead to desired structures, it is important to consider the mechanism responsible for the effective term. In what follows we assume a standard type-I, II or III seesaw mechanism (for a brief review of the seesaw types and their connection with leptogenesis, see, e.g., Ref.~\cite{Davidson:2008bu}).

In the type-I (type-III) seesaw, heavy right-handed fermion singlets (triplets) are added to the SM. The type-I seesaw Lagrangian for the neutrino sector becomes
\begin{equation}
\mathcal{L}_\nu=Y_D^{ij}\,\overline{L}_iH N_{j}+M_{R\,ij}\,\overline{N}^c_{i}N_{j},
\end{equation}
where $N_i$ are the right-handed neutrino fields, $Y_D$ is the Dirac-neutrino Yukawa coupling matrix, and $M_R$ is the heavy Majorana neutrino mass matrix. The type-III seesaw Lagrangian is similar, with the right-handed neutrino $N$ replaced by the fermion triplet, and the $SU(2)$ contractions appropriately changed.

Let us consider an arbitrary mass-independent leptonic mixing texture and assume three light neutrinos. Following the arguments of Ref.~\cite{Felipe:2009rr}, we can then state the following: If the matrix $M_R$ has the same parametrization as the light neutrino mass matrix $m_\nu$, i.e.,
\begin{equation}\label{MR}
M_R=a_R C+b_R P+c_R D \,,
\end{equation}
with $C, P$, and $D$ defined in Eqs.~\eqref{CPDstruct}, then the matrix $m_D=v Y_D$, where $v$ is the SM Higgs VEV, is also parametrized in the same way,
\begin{equation}\label{mD}
m_D = a_D C+ b_D P+ c_D D \,,
\end{equation}
and is thus symmetric. As discussed in detail in Sec.~\ref{ITB}, the assumption about the texture of the matrix $M_R$ given in Eq.~\eqref{MR} is justified in the context of family models with a group structure.

In the type-II seesaw framework, heavy scalar triplets $\Delta_a$ are added to the model, and the Lagrangian in the neutrino sector becomes
\begin{equation}
\mathcal{L}_{\nu}=Y_a^{ij}\overline{L}^c_iL_j\Delta_a \,.
\end{equation}
The neutrino mass matrix structure arising from these terms is controlled by the allowed contractions.

\subsection{Relevant invariants for TB mixing \label{ITB}}

In order to obtain the TB mixing from the invariants, it is necessary to discuss first the representations of the charged leptons and the mechanism responsible for the generation of the effective neutrino mass term. Within type-I seesaw, it is possible to draw some conclusions about the matrix $M_R$, since it is constructed from invariants with repeated representations. If $N$ belongs to singlet representations, there are too many parameters, and mass-independent textures cannot be generated without fine-tuning. Thus, $N$ must be a family triplet, and the invariant contractions can either be $NN$ (leading to $P$) or $NN\phi$ (leading to $C$ and/or $D$, depending on the group and the VEVs of the $\phi$ fields). The form given in Eq.~(\ref{MR}) is obtained, and, as stated after Eq.~(\ref{mD}), the Dirac-neutrino mass matrix $m_D$ needs to be symmetric. Furthermore, if we attempt to construct the symmetric contribution, due to the magic matrix $C$ in $m_D$, from a $LN\phi\sim\mathbf{1}^{3\varphi}_{0(0)} (\text{sym})$ invariant of $\Delta(3 n^2)$, the undesired antisymmetric contribution $\mathbf{1}^{3\varphi}_{0(0)} (\text{asym})$ is also generated, thus spoiling the TB mixing. Therefore, we must forbid both contributions by setting $a_D=0$ in Eq.~(\ref{mD}).

Within type-II seesaw, the effective neutrino mass matrix is obtained directly from repeated representations: $L$ must be a triplet representation with family invariant contractions $LL$ or $LL \phi$. Notice also that, although, in general, the SM Higgs $H$ and the $SU(2)$ triplet scalar $\Delta$ are considered as family symmetry singlets, and the triplet fields $\phi$ are added to the theory, it is possible to replace $\phi$ by assigning the $SU(2)$ multiplets $H$ and $\Delta$ to family triplets. By doing so, the theory becomes renormalizable.

Considering in detail the representations, we can also formulate general arguments to justify the absence or suppression of the democratic contribution to $m_\text{TB}$. This has been already discussed for $A_4=\Delta(12)$ models. Here, we generalize it for $\Delta(3 n^2)$ with $n>2$.

Consider, for instance, $n=3$, i.e., the $\Delta(27)$ group. There are only 2 triplet irreps, $(0,1)$ and $(0,2)$. In this case, the $LL$ term is not allowed and the two choices for three-triplet invariants, $(0,1) \times (0,1) \times (0,1)$ or $(0,2) \times (0,2) \times (0,2)$, are equivalent and always result in three simultaneous invariants with $q=0, 1$, and $2$.

Assume now $n=4$, i.e., $\Delta(48)$. This group has 5 triplet irreps which can be labeled as $(0,1)$, $(0,2)$, $(0,3)$, $(1,1)$, and $(3,3)$. The three-triplet invariant with $p=q=0$ can result e.g. from one of the outcomes of the product $(0,2) \times (0,1) \times (0,1)$. This invariant is never available with a repeated $(0,2)$ irrep, as required by the invariant $L L$. If we assign $L$ to $(0,1)$, though, it is possible to obtain simultaneously all three structures at the cost of an extra field: when the $(0,3)$ scalar aligns in the $(1,1,1)$ direction, the product $(0,3)\times(0,1)\times(0,1)$ allows the $q=1,2$ invariant necessary for the $C$ matrix, while the product $(0,2)\times(0,1)\times(0,1)$ allows the $q=0$ invariant from which both $P$ and $D$ can be constructed, as soon as the scalars align in the $(1,0,0)$ and $(1,1,1)$ directions, respectively. Within this context, $\Delta(48)$ is the smallest $\Delta(3n^2)$ group for which the effective neutrino mass matrix naturally contains the democratic structure and thus allows for an inverted light-neutrino mass spectrum.

\subsection{Implications for leptogenesis}

Type-I and type-III seesaw flavor models that lead to an exact mass-independent leptonic mixing have a vanishing leptogenesis $CP$ asymmetry in leading order~\cite{Bertuzzo:2009im,AristizabalSierra:2009ex,Felipe:2009rr}. This result does not necessarily hold in type-II seesaw models. In the latter case, the leptonic asymmetry is, in general, nonvanishing and, as it turns out, can be related to the democratic component of the neutrino mass matrix.

We recall that the unflavored leptonic $CP$ asymmetry $\epsilon_a$ in a type-II seesaw framework is proportional to
\begin{equation}\label{unflavoured}
\epsilon_a\propto
\text{Im}\left[\mu_a^\ast\mu_b\text{Tr}\left( Y^bY^{\dagger a}\right)\right],
\end{equation}
while the flavored asymmetries $\epsilon_a^{\alpha\beta}$ depend on
\begin{equation}\label{flavoured1}
\epsilon_a^{\alpha\beta}\propto
\text{Im}\left[\mu_a^\ast\mu_b Y_{\alpha\beta}^bY_{\alpha\beta}^{\ast
a}\right]
\end{equation}
and
\begin{equation}\label{flavoured2}
\epsilon_a^{\alpha\beta}\propto
\text{Im}\left[\text{Tr}\left(Y^{b\dagger}Y^a\right)
Y_{\alpha\beta}^bY_{\alpha\beta}^{\ast a}\right].
\end{equation}
In the above expressions, $\alpha$ and $\beta$ are flavor indices, and $\mu_a$ and $Y_a$ are the couplings of the scalar triplets $\Delta_a$ to the SM Higgs and charged lepton fields, respectively.

It is convenient to distinguish two cases, depending on whether $\Delta$ is a singlet or a triplet under the family symmetry. For leptogenesis to be viable, at least two scalar $SU(2)$ triplets are needed. If both are family singlets, then one of them can be associated to the $P$ contribution, and the other one to the $C$ contribution, in Eq.~(\ref{mnu3}). If a third scalar triplet is available, it may be associated to the democratic component $D$. In this minimal setup, unless a democratic contribution is present, the unflavored asymmetry \eqref{unflavoured} is zero, because the product of the $C$ and $P$ matrices is traceless\footnote{Notice however that, if each scalar triplet is simultaneously associated to the magic and $\mu$-$\tau$ symmetric contributions, the unflavored asymmetry \eqref{unflavoured} is, in general, nonvanishing.}. On the other hand, the flavored leptogenesis asymmetries do not necessarily vanish, even when the democratic component is absent. In the latter case, the combination in Eq.~\eqref{flavoured2} would vanish, but the one in Eq.~\eqref{flavoured1} would be, in general, nonzero.

If there are $\Delta$ family triplets, there must be at least one extra singlet or triplet. Otherwise, it is not possible to generate a mass-independent mixing in agreement with low-energy neutrino data. It can be shown that any contributions to Eqs.~\eqref{flavoured1} and \eqref{flavoured2} that involve components of the same triplet cancel out, but a nonvanishing asymmetry can result from the interaction of a given component of the triplet with the extra singlet or triplet. However, even in this case, the interaction of the triplet with the extra field will produce a traceless product in Eq.~\eqref{unflavoured}, unless one of the components is the democratic one (it should be noted that this requires an enlarged $\Delta$ field content, since two family triplets are necessary, plus either a third triplet or a singlet field). Furthermore, if $\Delta$ is a family triplet, there are additional complications in constructing the interaction term $\Delta H H$, which makes the choice of $\Delta$ as singlets more appealing.

\section{Other mass-independent structures}

In Sec.~\ref{OMI}, we have expressed a mass-independent mixing $U_X$ in terms of the TB mixing through Eq.~\eqref{KX}. In this way, we obtain a simple recipe to transform an existing TB model to another mass-independent mixing scheme. The process requires an appropriate change of basis in the group generators and also the rotation of the VEVs. The simplest way to obtain the rotated VEVs is by means of the transformation given in Eq.~\eqref{UW}. We write
\begin{equation}
 \langle \varphi \rangle = K_X
U(\omega)^\dagger \langle \varphi_\text{old} \rangle.
\end{equation}
This is equivalent to having rotated the basis as in Eq.~\eqref{KX} and solving for the VEV that leads to the desired mass matrix structure. Note that taking $K = \openone$ corresponds to TB mixing, and rotating the VEV $\langle \varphi_\text{old} \rangle = (1,0,0)^T$ leads to $\langle \varphi \rangle = (1,1,1)^T$ as expected.

As an explicit example, we describe next how to obtain from the $A_4$ symmetry the GR1 mixing scheme. The necessary VEVs do not appear natural and may be difficult to obtain from the group structure. Existing models that produce the GR1 mixing from a discrete group such as $A_5$ also appear to require involved VEVs and irreps with higher dimensions~\cite{Everett:2008et}. The $1'$ and $1''$ invariants in the rotated basis are not convenient to obtain a diagonal charged-lepton mass matrix for the GR1 model, but we can introduce additional fields with different VEVs in order to use the trivial singlet contraction. In order to obtain the non-TB mass invariant structure, we need intricate VEVs and extra fields to avoid the complicated $1'$ and $1''$ invariants. Therefore, it is fair to say that the $A_4$ group leads more naturally to TB mixing.

For the GR1 example, we assign the SM particles as $L \sim (0,1)$; $e_R,\,\mu_R,\,\tau_R\sim 1$; and $H\sim 1$. We introduce extra scalar fields $\phi_e,\,\phi_\mu,\, \phi_\tau$, and $\phi$ that are triplets under $A_4$ and are responsible for its spontaneous breaking. The Lagrangian is
\begin{align}\label{LX}
\mathcal{L}_{X} &=\frac{y_e}{\Lambda}\,\phi_e\left(LHe_R\right)+\frac{y_\mu}{
\Lambda }\,\phi_\mu\left(LH\mu_R\right)+\frac{y_\tau}{\Lambda}\,
\phi_\tau\left(LH\tau_R\right)\nonumber\\
&+\frac{y_\nu}{\Lambda}\,\left(LHLH\right)+\frac{y'_\nu}{\Lambda^2}\,
\phi\left(LHLH\right) \,,
\end{align}
where $\Lambda$ is the cutoff energy scale.
Note that the Lagrangian in the neutrino sector is equal to the one that generates TB mixing, but the charged lepton sector has been modified. The required scalar VEVs are
\begin{align}
\begin{split}
&\langle \phi_e \rangle=(1,0,0) \,,\quad \langle \phi_\mu
\rangle =(0,0,1)\,,\\
& \langle \phi_\tau \rangle =(0,1,0)\,,\quad
\langle \phi \rangle =(\sqrt{4-2\Phi},1,-1) \,.
\end{split}
\end{align}
If the rotated democratic contribution is absent, the predictions described in Sec.~\ref{Pheno} are preserved.

The Lagrangian in Eq.~\eqref{LX} can be used to predict other mixing structures with different VEV alignments. For instance, the GR2 mixing is obtained assigning $\langle \phi \rangle = (\sqrt{14-8\Phi},1,-1)$, while the bimaximal mixing corresponds to $\langle \phi \rangle = (-\sqrt{2},1,1)$. On the other hand, assigning the VEVs
$\langle \phi_e \rangle =(1,0,0)$, $\langle \phi_\mu \rangle=(0,1,2\sqrt{2})$, $\langle \phi_\tau \rangle =(0,-2\sqrt{2},1)$, and $\langle \phi \rangle =(1,1/\sqrt{3},\sqrt{2/3})$, we can predict the transposed TB mixing. Finally, the hexagonal mixing can be obtained with the enticingly simple VEV $\langle \phi \rangle =(\sqrt{2/3},1,1)$, so that a complete model for hexagonal mixing may be constructed by simply adding the appropriate aligning terms to the scalar potential.

\section{Conclusions}

We have shown that it is useful to decompose the effective neutrino mass matrix associated to the tribimaximal leptonic mixing into three independent parts [see Eq.~\eqref{mnu3}], including a democratic contribution. We concluded that this contribution controls interesting phenomenological consequences, such as the type of mass spectrum. In particular, we pointed out that the existence of a democratic contribution is necessary to obtain an inverted neutrino hierarchy. We then generalized those considerations to other mass-independent schemes, which are related to the TB form by a unitary transformation. We derived the same phenomenological results in terms of the rotated contribution of the given mass-independent structure.

We have also considered, in a model-independent context, the consequences of using different groups and choices of representations. We discovered that, for $\Delta(3n^2)$ groups producing TB mixing, the order $n$ of the group is very important and determines how natural it is to obtain a significant democratic contribution in the neutrino mass matrix. It would be interesting to extend the analysis based on the family symmetry invariants to other groups, such as $\Delta (6 n^2)$, which includes $S_4 = \Delta(24)$.

From a model-building viewpoint, we have seen that the type of seesaw mechanism responsible for neutrino masses plays an important role, as it restricts the viable choices of family representations. It also has important consequences for leptogenesis. While in minimal type-I and type-III seesaw scenarios that lead to a mass-independent leptonic mixing the leptonic $CP$ asymmetries are zero in leading order, in a type-II framework, leptogenesis is, in general, viable. We also concluded that the democratic contribution plays a relevant role in allowing unflavored type-II leptogenesis to proceed.

Finally, we have extended our analysis of TB mixing to other mass-independent mixing schemes and have shown how a given TB model can be modified to yield such structures. This process has been illustrated by constructing a toy model that predicts the golden ratio mixing, which, although not as elegant as the respective TB model, is simple when compared to other models that predict the same mixing. The necessary VEV alignment to obtain the golden ratio mixing has been presented, and, for completeness, we have also given the alignments that would lead to the bimaximal and transposed tribimaximal mixing patterns, as well as the relatively simple alignment needed for the hexagonal mixing. We have then concluded that, within this kind of rotated model, the special connection between the would-be democratic contribution and phenomenology is preserved.

\section*{Acknowledgements}

The work of I.d.M.V. was supported by Funda\c{c}\~{a}o para a Ci\^{e}ncia e a Tecnologia (FCT, Portugal) under Grant No. SFRH/BPD/35919/2007 and by DFG Grant No. PA 803/6-1. The work of H.S. was supported  by FCT under Grant No. SFRH/BD/36994/2007. This work was partially supported by FCT through the projects CFTP-FCT UNIT 777,  PTDC/FIS/098188/2008, and CERN/FP/109305/2009, which are partially funded through POCTI (FEDER).

\end{document}